\documentclass[aps,prl,twocolumn,showpacs,longbibliography,floatfix]{revtex4}
\usepackage{graphicx}
\usepackage{amsmath}
\usepackage{amssymb}
\usepackage{color}
\usepackage[normalem]{ulem}
\usepackage{epsfig}
\graphicspath{ {figures/} }

\begin{document}
\title{Neutral dynamics with environmental noise: age-size statistics and species lifetimes.}

\author{David Kessler$^{1}$, Samir Suweis$^{2}$, Marco
Formentin$^{3}$  and Nadav M. Shnerb$^{1}$}

\affiliation{$^{1}$Department of Physics, Bar-Ilan University,
Ramat-Gan IL52900, Israel.\\  $^{2}$Physics and Astronomy Department
`G. Galilei' \& CNISM, INFN, University of Padova, Via Marzolo 8,
35131 Padova, Italy. \\  $^{3}$UTIA, Czech Academy of Sciences, Pod
Vod\'{a}renskou v\v{e}\v{z}\'{i} 4, 18208 Prague, Czech Republic}

\begin{abstract}
Neutral dynamics, where taxa are assumed to be demographically
equivalent and their abundance is governed solely by the
stochasticity of the underlying birth-death process, has proved
itself as an important minimal model that accounts for many
empirical datasets in genetics and ecology. However, the restriction
of the model to demographic [${\cal{O}} ({\sqrt N})$] noise yields
relatively slow dynamics that appears to be in conflict with both
short-term and long-term characteristics of the observed systems.
Here we analyze two of these problems - age size relationships and
species extinction time - in the framework of a neutral theory with
both demographic and \emph{environmental} stochasticity. It turns
out that environmentally induced variations of the demographic rates
control the long-term dynamics and modify dramatically the
predictions of the neutral theory with demographic noise only,
yielding much better agreement with empirical data. We consider two
prototypes of "zero mean" environmental noise, one which is balanced
with regard to the arithmetic abundance, another balanced in the
logarithmic (fitness) space, study their species lifetime statistics
and discuss their relevance to realistic models of community
dynamics.
 \end{abstract}
\maketitle

\section{Introduction}

The theory of population and community dynamics is a central
mathematical tool in many branches of life science, including
genetics, ecology and evolution. This broad field of research is
dominated by two competing and complementary paradigms. Darwinian
natural selection suggests that the fitness of phenotypes and
species is different and stresses deterministic effects like the
survival of the fittest, downplaying the role of noise and
fluctuations. On the other hand, neutral theories assume that
selective effects are relatively weak and different taxa/haplotypes
admit almost identical fitness, so that the main driver of
population dynamics is stochasticity.

Within the neutral framework, first Kimura's theory of molecular
evolution \cite{kimura1984neutral} and more recently Hubbell's
universal neutral theory of biodiversity (UNTB)
\cite{hubbell_book,muneepeerakul2008Nature,TREE2011} have both
attracted a lot of attention. In the latter case, the successful
explanation of empirical species abundance distribution curves by a
simple theory with only two parameters
\cite{maritan1,marquet2014theory} appears as a very appealing
minimalistic model, especially when compared to niche-based
approaches that usually require the reconstruction of many
parameters (such as the relative fitness of species), a very
difficult task in high-diversity assemblages
\cite{connolly2014commonness}.

These neutral models  assume that the main driver of community
dynamics is demographic stochasticity, i.e., the noise embodied in
the birth-death process of individual agents, with (if the whole
community has to keep a fixed size) the expected number of
descendants for each individual being precisely one.
 Accordingly,  a population
of $N$ individuals will produce $N$ offspring on average, and the
per-generation fluctuations will be proportional to $\sqrt{N}$,
which is quite a weak noise in the limit of large $N$. This
restriction of the neutral model to pure demographic noise leads to
a few severe difficulties when its predictions are compared with
empirical patterns.

On evolutionary time scales, the two main unsolved  problems are the
age-size relationships and species extinction time. In a neutral
theory with pure demographic noise both the age (measured in
generations) of a species and its time to extinction are
proportional to its abundance. This timescale is ridiculously long
for, inter alia, various species of trees (with generation time of
50y) and for passerine birds (generation time 3y), as noted by many
authors
\cite{nee2005neutral,ricklefs2006unified,allen2007setting,chisholm2014species}.
The fossil data, which indicate that species lifetime is typically a
few million years, is again in contradiction to the $N$ generation
estimate for common species. As Robert Ricklefs summarized his
findings \cite{ricklefs2006unified}, ``drift is simply too slow to
account for the rate of turnover of passerine birds''. The idea of
protracted speciation \cite{rosindell2010protracted}, suggested to
account for the apparent underrepresentation of rare species and
their relatively long lifetime, cannot resolve these difficulties.

On the ecological time scale, empirically observed fluctuations in
abundance are usually too strong to be explained by UNTB
\cite{leigh2007neutral,feeley2011directional,Kalyuzhny2014niche,chisholm2014temporal}.
Moreover, UNTB cannot explain the scaling of fluctuations variance
with population size: with pure demographic noise the theory
predicts a linear scaling but empirical analyses show a prevalence
of super-linear dependence \cite{Kalyuzhny2014temporal}. The decay
of community compositional similarity is again much faster than the
UNTB predictions \cite{kalyuzhny2014generalized}.

A very plausible (and one in any case necessitated by biological
reality) generalization of the neutral theory that may resolve many
of these problems while preserving the minimalistic character of the
model, is to add environmental stochasticity to the dynamics. In
nature the demographic response of individuals to environmental
variations is (at least partially) correlated within species, while
different species may show transient fitness advantage at various
times due to differences in their temporal niches. In our proposed
generalized neutral model all individuals are demographically equal
\emph{on average} but the relative fitness of a population
fluctuates in time. Environmental stochasticity generates ${\cal
O}(N)$ short-term fluctuations in population abundance, closer in
size to those observed in reality.

\begin{figure*}
\begin{center}
\includegraphics[width=19cm]{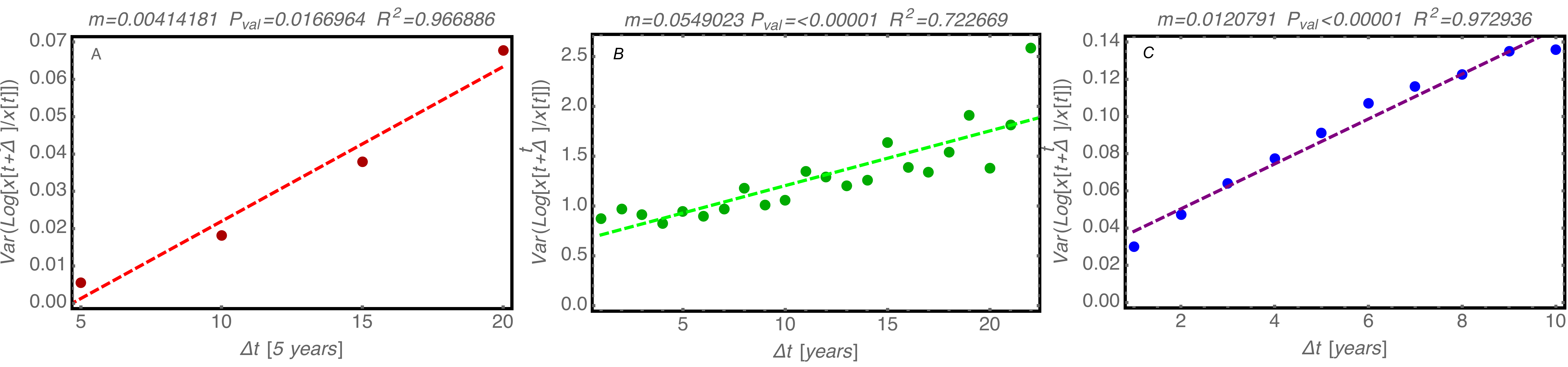}
\vspace{-0.cm}
\end{center}
\caption{Variance through time plots for three empirical datasets:
A) Trees of the tropical forest in Barro Colorado Island (BCI),
considering trees with diameter $>10 mm$ (data are from
www.ctfs.si.edu) from 1990 to 2005 (every 5 years).  B) Fish samples
collected from the cooling -water filter screens at Hinkley Point B
Power Station in Bristol, from 1980 to 2008
\cite{magurran2003explaining}. We only consider estuarine fish
species (not crustaceous organisms). C) herbaceous plant dataset
\cite{adler2007long} comprising a series of 51 quadrats of 1 m$^2$
from mixed Kansas grass prairies where all individual plants were
mapped every year from 1932 to 1972. In all cases we gathered
$log(N_{t+\Delta t}/N_t)$ from every two same species abundance
observations  that are $\Delta t$ apart, and plotted the variance of
this quantity versus $\Delta t$ (the details of this technique are
presented in \cite{Kalyuzhny2014niche}). $m$ denotes the angular
coefficient of the linear fit of data points, while $P_{val}$ and
$R^2$ give the significance and the goodness of the fit,
respectively. In all three cases the variance appears to grow
linearly in time, as suggested for a system with logarithmically
balanced environmental noise, where the slope $m$ indicates the
value of the effective $D$ defined in (\ref{dd}). The intercept of
the linear fit is above zero for the fish and the grassland,
indicating the effect of sampling errors \cite{Kalyuzhny2014niche}.
The negative intercept for trees indicates a delayed response of the
system to the changing environment. Since the timescales are
relatively small, the same analysis for arithmetically balanced
noise, where the variance of $(N_{t+\Delta t} - N_t)/N_t$ is plotted
against $t$, yields almost the same results (not shown).}
\label{fig1r}
\end{figure*}

To provide an order of magnitude estimate for the strength of
environmental noise in empirical systems, the variance through time
plot of the logarithmic abundance ratio is presented in Fig.
\ref{fig1r} for three empirical datasets, showing the linear
increase characteristic of environmental stochasticity. As discussed
in \cite{Kalyuzhny2014niche}, the slope of this curve is
proportional to the strength of environmental variations. In the
discussion section we plug this number into our mathematical
expressions to show that realistic noise indeed solves the UNTB
timescale problem.

Our first major result, then, is that a neutral model with both
demographic and environmental noise may solve the mismatch between
the predictions of the purely demographic UNTB and the empirical
evidence, shortening significantly the number of generations needed
for the ancestry of a single founder to reach large abundance. The
linear scaling of time (in generations) with the abundance $N$ is
replaced by logarithmic scaling, yielding reasonable age-size
estimations. Our second result is that the time to extinction of an
abundant species is shortened significantly. As described below, the
various possibilities obtained for species lifetime statistics
correspond to previously, evidence-based, suggestions.

The implementation of environmental stochasticity into a neutral
model poses an important conceptual problem. As noticed a while ago
\cite{lewontin1969population} there are two different scenarios for
balanced (zero mean) environmental stochasticity: the balance may be
either arithmetic (i.e., the noise statistics is such that the mean
abundance is kept fixed; for example if during a bad year the
population shrinks by $1/2$, during a good year it grows by 50\%) or
logarithmic (in a bad year it shrinks to 1/2, say, and in a good
year the population doubles). These two scenarios are analogous to
the use of Ito or Stratonovich calculus in the white noise limit
\cite{kupferman2004ito,suweis2011prescription}.

For a fixed-size community, where different species are playing,
more or less, a zero sum game, none of these scenarios provide a
satisfactory description of the dynamics. Under arithmetically
balanced noise all species species eventually go extinct with
probability one, while under log-balanced noise the size of the
community grows exponentially, both in contrast with the fixed size
requirement. Introducing environmental stochasticity into the UNTB
is a delicate task, in which the interplay between environmental and
demographic fluctuations and other effects should be taken into
account. Here we do not solve this problem, although we provide some
preliminary considerations in the discussion section.

However, any model with balanced noise must be somewhere  between
the log-balanced and the arithmetic-balanced extremes considered
here and, as will be shown below, in \emph{both} cases the
abundance-age relationships are logarithmic so the species lifetime
problem is solved in any case. On the other hand, the log and the
arithmetic dynamics differ dramatically with respect to species
lifetime: the $t^{-2}$ tail of lifetime statistics predicted by the
purely demographic theory \cite{pigolotti2005} is replaced by a
slower ($t^{-3/2}$) decay if the noise is log balanced and by much
faster, exponential decay, for arithmetically balanced
stochasticity. In the literature one may find empirical evidence for
all these behaviors \cite{bertuzzo2011PNAS,Suweis2012JTB}; the
combination of demographic and environmental stochasticity is
endowed with the required flexibility to account, in the appropriate
regime of parameters, for many observed patterns of species/genus
lifetime statistics.

\section{Theoretical Framework}

Demographic noise reflects variability in reproductive success which
is uncorrelated among individuals in the population. This noise is
fully characterized by $P_n$, the probability of an individual to
produce $n$ offspring during its lifetime. The average number of
offspring is $\sum_n n P_n = R_o$, and for a fixed size population
$R_o = 1$. Another important parameter is the variance of the number
of offspring $\sigma^2$, which measures the differences in
reproductive success among individuals: in animal and human
populations, where the number of offspring is between zero and 10,
say, and $P_n$ decays sharply with $n$, the variance is of order
unity; for a virus population, where some viruses infect a cell and
produce 10000 offspring while the others die childless, the variance
is much larger.

The theory of a fixed-size population under pure demographic noise
traces back to the work of Galton and Watson \cite{galton1874}. In
Appendix  A we present a generating function analysis of this case.
Species lifetime statistics are reflected in $\delta(t)$, the chance
of the lineage of a single individual to survive until $t$ (time is
measured in units of generations), which is shown to satisfy
\begin{equation} \label{eqn1}
\frac{d\delta}{dt} \sim -\frac{\sigma^2}{2} \delta^2(t).
\end{equation}
This implies that the survival probability of a species decays at
long times like $1/t$ and that the statistics of species lifetime
admits a $1/t^2$ tail \cite{pigolotti2005}. Since the average
abundance of a species is constant, the fact that only a fraction
$2/(\sigma^2 t)$ of the species survive by time $t$  implies that
their mean abundance should grow linearly with t,
\begin{equation} \label{eqn2}
N_{surv}(t) \sim \frac{\sigma^2}{2} t.
\end{equation}
where $N_{surv}(t)$  is the average population abundance at $t$,
\emph{conditioned on survival}. These results demonstrate the two
main problems of a neutral theory with pure demographic noise, when
confronted with empirical findings and a priori considerations:
first, the time needed for a species to reach abundance $N$
(starting with one individual, i.e., point speciation) is typically
of order $N$ generations. Second, the typical time to extinction of
a species of abundance $N$ is again of order $N$ generations, since
the theory is neutral and the ancestry of any individual evolves
independently (Formally the average time to extinction diverges due
to the $1/t^2$ tail). Slight modification of the model, like
implementing a zero-sum game in the community (such that the size of
the community is kept fixed in the strong sense, as opposed to
keeping the average size fixed by assuming $R_o=1$), do not
significantly change
 these conclusions \cite{chisholm2014species}.

Adding \emph{environmental stochasticity} to this model implies
that, as the environmental conditions vary, the demographic success
of the whole population varies accordingly, so the average
reproductive success, $R_o$, becomes time dependent. In some cases,
discussed in \cite{kalyuzhny2014generalized}, the addition of
environmental noise leads to a \emph{stabilization} of the
populations around some equilibrium value because of the (quite
counterintuitive) \emph{storage effect}. This paper deals with
models that have no such effect, a set which includes  any
environmental noise generalization of the dynamics considered in
\cite{maritan1}.

Environmental conditions admit some correlation time $T$. Speaking
about a good/bad year (in terms of precipitation, winds etc.) one
assumes that the environment (in the general sense, including the
effects of competition with other species) was, in general,
favorable or hostile to a specific species during this period. If
the demographic rates are kept fixed during $T$ and the abundance of
a certain species at time $s$ is $N_s$, then typically $N_{s+1} =
e^{\gamma_s} N_s$ where $\gamma_s = (R_o -1) T$. The simplest way to
define a ''balanced" environmental noise is to assume that
$\gamma_s$ (the fitness parameter, or the deviation of $R_o$ from
unity) is an identically distributed random variable with zero mean
and variance
\begin{equation} \label{dd}
{\overline{\gamma_s^2}} \equiv 2D, \end{equation} where the overbar
denotes an average. In this scenario the steps are balanced in the
logarithmic space (the expectation value of $log(N)$ is kept fixed)
but the arithmetic mean of $N$ is growing in time, since
$\overline{exp(\gamma)}>1$. The reason for this is the asymmetry
between growth and decline: since the per capita growth/decay rate
is kept fixed for some time, the overall demographic benefit for the
population includes not only the birth originated from the
individuals that were present at $s$, but also from the individuals
that were born between $s$ and $s+1$, and the opposite is true in
the case of decline. The response of the population to varying
environmental conditions has an ``inertia'' that increases the
overall demographic benefit during good times relative to the loss
suffered during bad times, hence producing a net bias towards
growth.

However, in many scenarios the per-capita growth rate decreases when
the population increases. For example, in a fixed size community
with a zero sum game like the one considered in
\cite{kessler2014neutral}, the fitness of a species determines its
chance to replace an individual of another species by its own
offspring; the more abundant a species is, the greater the chance of
intraspecific competition so the inertia is weaker. In its extreme
limit one can model this kind of behavior by taking $N_{s+1} =
\gamma_s N_s$, where $\gamma_s$ is again a balanced noise. Now it is
the expectation value of $N$ (rather than $log(N)$) which is kept
fixed. We define these two types of environmental noise as
logarithmically balanced (case A, where the "opposite" of $N \to
N/2$ is $N \to 2N$) and arithmetically balanced (case B, the
"opposite" of $N \to N/2$ is $N \to 3N/2$).  Realistic systems with
environmental stochasticity are, most likely, somewhere between
these two extremes, but the solutions we present below for these two
scenarios provide the basic insights needed for consideration of the
generic case, as explained in the discussion section.

\section{Age-abundance relationships and lifetime statistics}

Within the above theoretical framework, we have developed and solved
a model of population dynamics under both environmental and
demographic noise, making a distinction between logarithmically
balanced (case A) and arithmetically balanced (case B)
stochasticity. The results have been derived for a model with a
geometric distribution of offspring, but, as explained in Appendix
C, they are valid for any realization of the demographic noise
provided that one is interested in the long-term behavior of the
system.

\subsection{Survival probability}

First we consider the survival probability, assuming a single
individual (point speciation) at $t=0$. It turns out (see Appendices
B and C) that the survival probability at long times is governed by
the equations:

\begin{eqnarray} \label{res}
 \frac{d \delta(t)}{dt} =  \gamma_t \delta(t)  + D \delta(t) - \frac{\sigma^2}{2} \delta^2(t) \qquad \rm{case \ A} \nonumber \\
  \frac{d \delta(t)}{dt} =  \gamma_t \delta(t)   - \frac{\sigma^2}{2}  \delta^2(t) \qquad \rm{case \ B}
\end{eqnarray}

In the limit  $\gamma = D = 0$ both equations reduce to
(\ref{eqn1}), the equation obtained previously for pure demographic
noise, as expected. Moreover, substituting $\delta =
2D\tilde{\delta}/\sigma^2 $, changing the time units with  $t =
\tau/D$ and rescaling the noise term to unit white noise, one finds
a $D$ independent equation. This implies that a plot of $\delta t
\sim \tilde{\delta} \tau$ vs. $Dt = \tau$ should be independent of
the strength of the environmental noise.

From Eq. (\ref{res}) one may extract the long-time asymptotics of
the survival probability. When the environmental stochasticity is
relatively weak the short time chance of extinction is controlled by
the demographic noise. This leads us to propose that, for both the
logarithmic and the arithmetic case,
\begin{equation}
 \delta(t) = \frac{{\cal F}(tD)}{t}.
\end{equation}
where the form of ${\cal F}(tD)$ is obtained analytically in
Appendix B II.

In \textbf{case A} (log balanced) ${\cal F}(x)$ approaches
$\sqrt{x/\pi}$ when $x \to \infty$ and unity
  when  $x \to 0$. Accordingly, the chance of extinction \emph{at} the $t$-th generation crosses over from the Galton-Watson universal
 limit (characterizing a process with pure demographic noise) $t^{-2}$ to the first passage time asymptotics $t^{-3/2}$
at $t^* \sim 1/D$.

The intuition behind this result is clear. In the absence of
demographic noise, the population preforms an unbiased random walk
in the logarithmic space, hence it will survive until it reaches a
threshold at, say, $log(N)=0$ (a single individual). The theory of
first-passage time for a 1d random walker tells us that at long time
the chance to survive decays like $t^{-1/2}$. Since populations that
stay alive for a long time typically reach high abundance, the long
time behavior is controlled by this term \cite{geiger2001survival},
while at shorter timescales environmental stochasticity is too weak
and the behavior is controlled by the $t^{-1}$ term of the purely
demographic process.

\begin{figure*}
\begin{center}
\includegraphics[width=15cm]{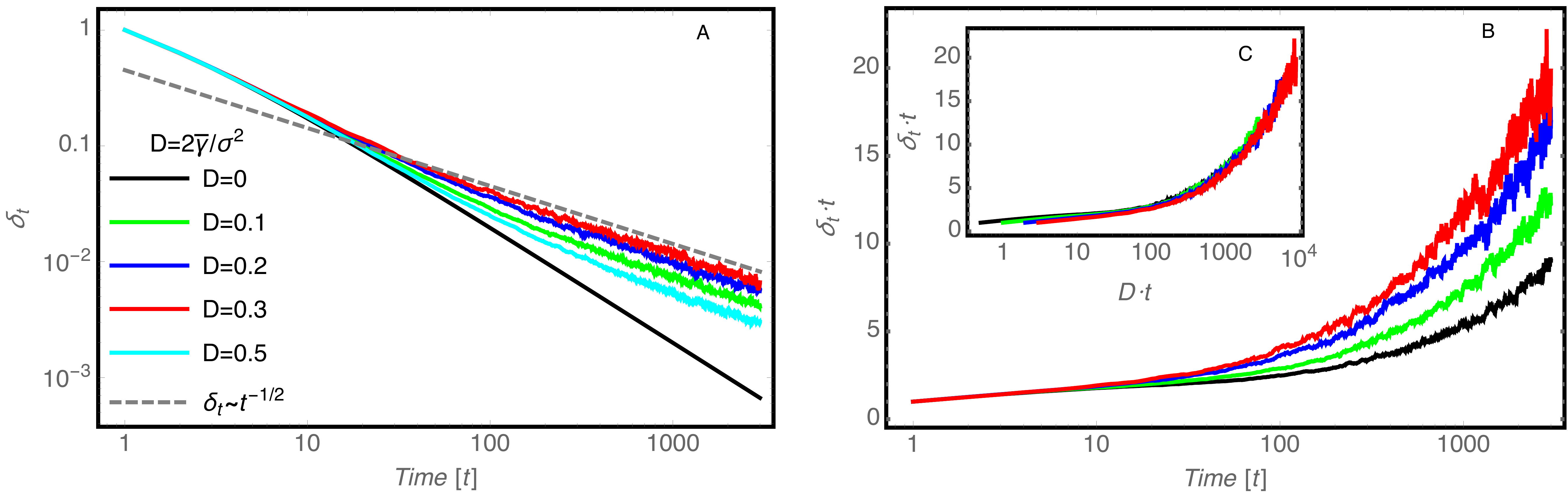}
\vspace{-0.cm}
\end{center}
\caption{A) A plot of $\delta_t$, the chance to survive at time $t$
for a certain levels of overall diffusion coefficient $D$ (log-log
scale). The simulated process is the standard Wright - Fisher
process with non - overlapping generations. B) Plotting $\delta_t
\cdot t$ versus time in linear-log scale versus time; C) A plot of
$\delta_t \cdot t$ against $t\cdot D$. In this case the data
collapses as predicted by Eq. (\ref{res}). This collapse suggests
that the proper scaling of the cumulative survival probability is
$\delta_t=\frac{\mathcal{F}(tD)}{t}$, with $\mathcal{F}(x) \sim
x^{1/2}$ for $x \rightarrow \infty$ and  $\mathcal{F}(x) \sim
const.$ when  $x \rightarrow 0$ }  \label{fig2}
\end{figure*}

 Figure \ref{fig2} shows simulation results for the  survival
  probability in case A, i.e., the chance that the system did not go extinct
  until $t$, $\delta(t)$. Indeed, this quantity decays like $t^{-1}$
  when the process is purely demographic, and it switches to
  $t^{-1/2}$ behavior at long times when the system is subject to environmental noise. Moreover, when $t\delta(t)$ is
  plotted against $Dt$ the data collapses  as predicted above.

\begin{figure}
\vspace{-3cm}
\begin{center}
\includegraphics[width=8cm]{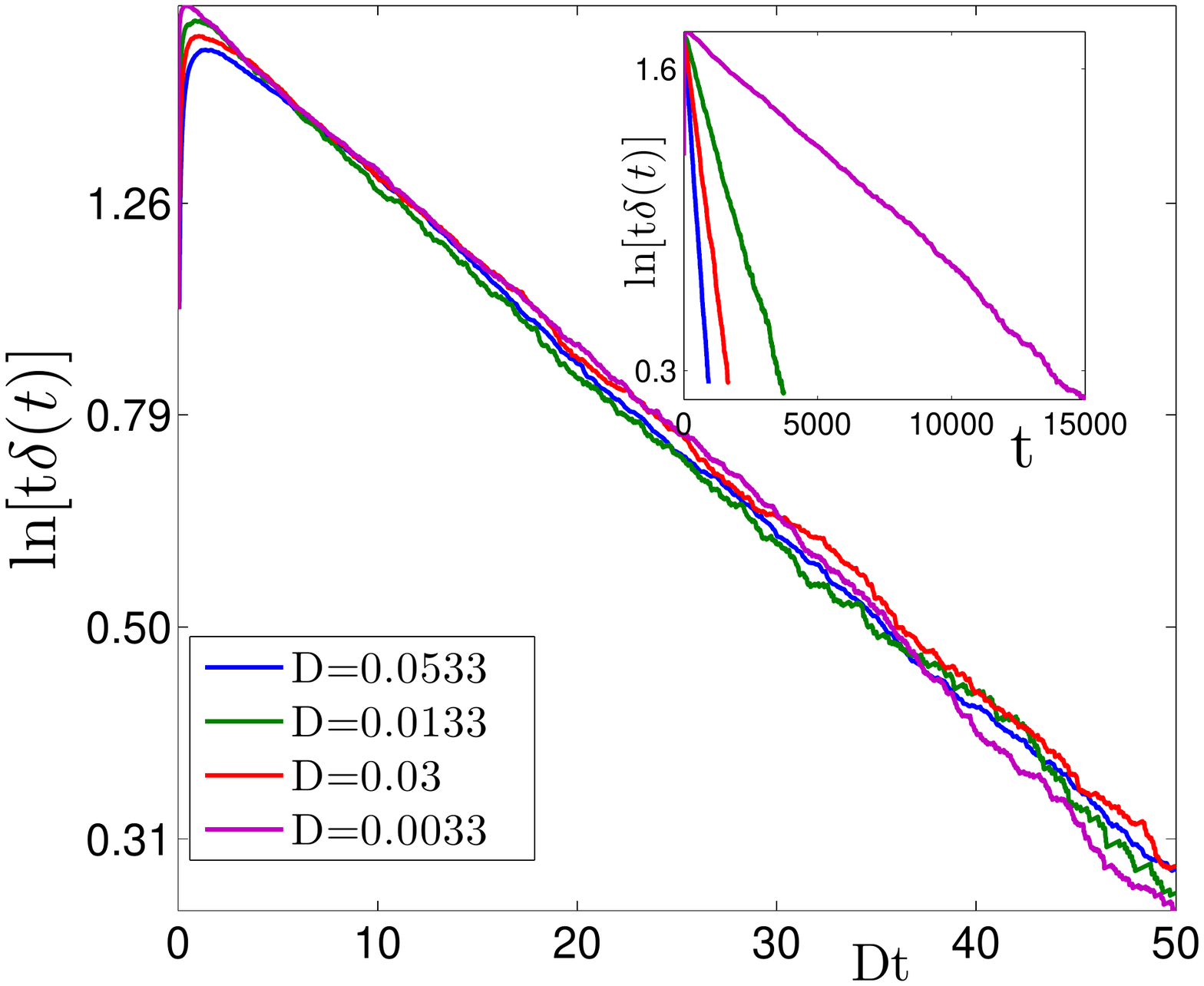}
\end{center}
\vspace{-3cm} \caption{The survival probability of a species,
$\delta$, vs. time in case B (arithmetically balanced noise). The
logarithm of $t \delta(t)$ is plotted against $t$ (inset), showing
an exponential decay with $D$ dependent slope. When plotted against
$Dt$ (main panel) the data collapse, indicating the dependence of
${\cal F}$ on $Dt$. } \label{fig4}
\end{figure}

In \textbf{case B} (arithmetic balance) ${\cal F}(x)$ approaches
$exp(-x/4)$ when $x \to \infty$ and unity when  $x \to 0$, as
demonstrated in Figure \ref{fig4}. Again the short-term behavior is
controlled by demographic noise, but the long-term survival
probability is no longer a power law. The reason is that, unlike the
log-balanced noise, in case B most of the species are shrinking in
time (as can be seen easily by tracing the 1/2-3/2 and 1/2-2
processes to the next generations).

Returning to the empirical species lifetime problem, for a neutral
theory with pure demographic noise the number of generations needed
for a population to shrink to zero is typically its abundance,
yielding unrealistic lifetimes for common species. When
environmental noise is introduced, the long-term dynamics (where
long is relative to the strength of the environmental noise, i.e.,
$t \sim 1/D$) is dominated by $x=Dt$, so a decent amount of
environmental stochasticity will shorten significantly the lifetime
and will solve the problem with ${\cal O}(N)$ generations scaling.
Regarding the statistics of the species lifetime distribution, all
the possibilities considered here - scaling with $t^{-3/2}$ in case
A, $t^{-2}$ for pure demographic noise and exponential decay - have
been suggested in the literature based on the analysis of fossil
data, see \cite{pigolotti2005} and references therein.

\subsection{Age-abundance relationships}

Now let us turn to the dependence of abundance on the species' age.
In Appendix B we define and calculate ${\bar \zeta(t)}$, the mean
abundance at time $t$ of the ancestry originated from a a single
individual at $t = 0$.
 However, when examining an empirical community one
considers only the species that have not yet gone extinct, so the
relevant quantity for the age-abundance relationships is the average
size of a species at $t$ conditioned on non-extinction, $N_{surv}(t)
= \zeta(t)/\delta(t)$. As shown in Appendix B-III, there are strong
differences between logarithmic and arithmetic noise, although the
bottom line is similar.

In \textbf{case A} (log balanced) $\zeta(t) \sim \frac{2}{D} \left(
e^{Dt/2}-1 \right)$. Since $\delta \sim t^{-1/2}$,
\begin{equation} \label{eq6}
N_{surv}(t) \sim \frac{2}{Dt^{1/2}}  e^{Dt/2}.
\end{equation}
Thus the abundance of the surviving species, instead of growing
linearly, is growing \emph{exponentially} in time. In Appendix B-IV
we explain that the average result presented here and the typical
result differ from each other due to the skewness of the
distribution, so in the \emph{typical} case the growth of $N_{surv}$
is subexponential
\begin{equation}\label{eq6t}
N_{surv}^{typ}(t) \sim e^{\sqrt{2Dt}}.
\end{equation}

For \textbf{case B}, on the other hand, $\zeta = 1$, since the noise
in balanced in the real space. On the other hand $\delta$ is
decaying exponentially, so the net result is again,
\begin{equation} \label{eq7}
N_{surv}(t)  \sim e^{Dt/4}.
\end{equation}

\section{Discussion and Conclusions}

The failure of the UNTB to account for dynamic patterns of
populations and communities is known for a long time
\cite{leigh2007neutral}, and was stressed recently by many authors.
Basically, the   ${\cal O}\sqrt{N}$ scaling of the demographic noise
makes it inadequate to account for the observed fluctuations on all
timescales. The tempo of the dynamics may be accelerated if one
assumes a very large value of $\sigma^2$ (as suggested, essentially,
in \cite{allen2007setting}, see \cite{chisholm2014species}) or by
keeping the generation time as a free parameter (see, e.g.,
\cite{azaele2006dynamical}), but any of these approaches carries its
own difficulties.

On the other hand, environmental stochasticity is known to be
ubiquitous in living systems, affecting communities even under the
most stable conditions (see, e.g., \cite{hekstra2012contingency}).
Incorporating this mechanism into the neutral theory is a required
step in any case \cite{loreau2008species}. We have already showed
that environmental noise increase substantially the heterogeneity of
the species abundance distribution (see \cite{kessler2014neutral},
Eq. 3), a feature that may account for the empirical results
analyzed in \cite{connolly2014commonness}.  The fact that this noise
increases temporal fluctuations and decreases the timescale makes
this project even more attractive.

Under environmental stochasticity, the demographic rates of all
individuals belonging to a species (roughly speaking, their fitness)
are fluctuating coherently in time, and the species abundance varies
accordingly. As explained, the environmental noise may be "neutral"
in two different senses. One scenario is when the relative fitness,
when averaged over time, will be zero, this corresponds to
logarithmically balanced noise or case A considered above. The other
scenario, case B, occurs when the time average of the demographic
gain is zero.

As shown above, these two cases correspond to two different species
lifetime statistics. The chance of a species to survive decays like
$1/t^{3/2}$ in case A and exponentially in case B. All these
possibilities, $-3/2$ law, $-2$ law and exponential, were suggested
for the tails of species lifetime distributions as extracted from
fossil data \cite{pigolotti2005}. More important is the transition
of the general scaling from $t$ (measured in units of generations)
to $Dt$, allowing the environmental noise to control the extinction
times.

\begin{figure}
\begin{center}
\includegraphics[width=8cm]{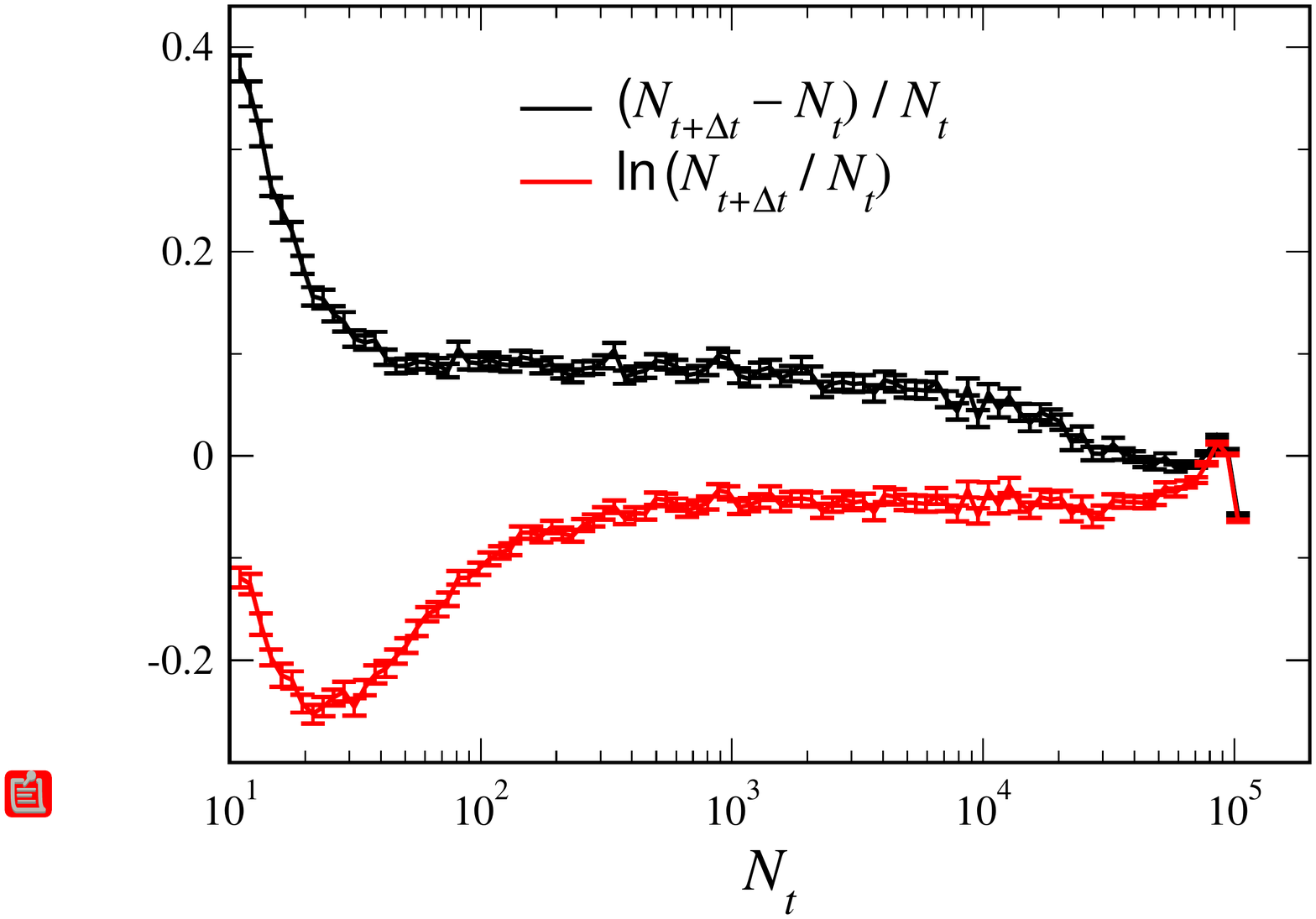}
\end{center}
\vspace{-0.8cm} \caption{The mean of the logarithmic (red) and
arithmetic (black) abundance fluctuations as a function of the
abundance, extracted from  simulation of a neutral community model
with both demographic and environmental noise. A Moran-type
continuous time process with discrete individuals was simulated. At
each small time step ($dt=0.1$) we generate the number of births and
deaths for each species.  The number of births is a Poisson
distributed number with mean $\alpha_i(t) dt$, where $\alpha_i(t)$
is the instantaneous growth rate (fitness) of species $i$. The
number of deaths in the species is binomially distributed, with the
probability $N_\mathrm{tot}/K$ per individual, where
$N_\mathrm{tot}$ is the instantaneous size of the community and
$K=10^5$. In addition, a Poisson number of new immigrants is drawn,
with mean $\mu dt$, where $\mu=2$.  Each immigrant founds a new
species. The birth rates $\alpha_i$ are given by $\alpha_i =
exp(\gamma_i)$ where the $\gamma_i$ are generated by an
Orenstein-Uhlenbeck process, $\dot{\gamma}_i = -\gamma_i/\tau +
\eta$, so that they are Gaussian distributed with mean 0 and
variance 0.001 and correlation time 2. This way the births rates are
guaranteed positive. } \label{fig5}
\end{figure}

For the average abundance of a species at time $t$ after point
speciation conditioned on non-extinction, $N_{surv}(t)$, we obtained
exponential growth (in case B) and a typical stretched exponential
growth (in case A) . This appears to solve the``age-size" problem
\cite{ricklefs2006unified,nee2005neutral}: while the time to the
most recent common ancestor scales with abundance in a purely
demographic neutral model, it scales with the logarithm (or
logarithm squared) of $N$ in the presence of demographic
stochasticity. Eqs.(\ref{eq6t}, \ref{eq7}) suggest that the time
from speciation to abundance $N$ scales like $\log(N)/D$ generations
(case B) or with $\log^2 N/(2D)$ (case A). If, for example, one
considers a set of $10^9$ conspecific trees for a frequent species
in the Amazon basin (this is close to the contemporary figure, see
recent survey in \cite{ter2013hyperdominance}), with about a 50y
generation time, the neutral theory suggests $50 \cdot 10^9$
generations, more than the age of the universe. On the other hand,
the left panel of Fig. \ref{fig1r} suggests that,measured in units
of a single generation (50y) $D \sim 0.1$, so the neutral theory
with environmental noise modifies the speciation time to about
50,000 years before present (case B) or about 150,000 ybp (case A).
This, of course, is an extrapolation since the environmental noise
may be either much smaller (if some of the short-term fluctuations
are averaged out due to balancing effects like the increase of
species specific parasites) or much larger (if extreme events does
not appear in the half-generation window considered), but the scales
are clearly small enough to solve the age-abundance problem even if
the estimation for $D$ is modified by an order of magnitude.

Finally, we would like to comment about the generic case.  When a
community is subject to a stabilizing mechanism that keep its size
fixed or almost fixed, the dynamics of a single species must be
somewhere between the two extremes considered along this paper. Pure
case A dynamics cannot hold as there average population increases
exponentially, while pure B dynamics is rejected since all species
go extinct with probability one so the size of the community must
shrink eventually to zero. In Figure \ref{fig5} we plotted the mean
of $(N_{t+\Delta t}-N_t)/N_t$ (positive in pure case A, vanishes in
case B) and the mean $ln(N_{t+\Delta t}/N_t)$ (vanishes in pure case
A, negative in case B) vs. $N$ as extracted from simulation of a
community model with environmental noise (species specific randomly
varying fitness). The situation is clearly between case A and case B
- the arithmetic mean is slightly positive, the log mean is slightly
negative - and shows some nontrivial $N$ dependence. Our approach
show how a rich behavior may arise in a simple model where we do not
model interaction among different species. Further works will
investigate the effect of environmental noise when  considering a
whole ecological community where species explicitly interact among
each other.



{\bf Acknowledgments} DAK acknowledges the support of the Israel
Science Foundation.  NMS acknowledges the support of the Israel
Science Foundation BIKURA grant no. $1026/11$. SS thanks the Physics
\& Astronomy Department, University of Padova for financial support
through the senior grant N. 13.32.2. MF acknowledges financial
support of GA\v{C}R grant P201/12/2613.

\bibliography{references}

\newpage

\onecolumngrid

\section{Appendix  A: Survival probability in a model with pure demographic noise}

We start by considering an event of point speciation, i.e., a new
taxon that appears as one individual, the founder. Let us denote the
chance that a single individual has $n$ descendants in the $s$th
generation by $P^{(s)}_n$. In a constant population, perfectly
neutral scenario with pure demographic noise, the number of
offspring, $n$, that every given individual produces is independent
of time  and identically distributed, given by
 $P^{(1)}_n$.  The average number of offspring is unity, i.e., $\sum_n n P^{(1)}_n
=1$  and the strength of the demographic noise is characterized by
$\sigma^2 \equiv \sum_n n(n-1) P^{(1)}_n$. This scenario was first
analyzed by Galton and Watson (GW), who showed that such a
population goes extinct with probability one \cite{galton1874}.

To  generalize the GW result, one defines the generating function
$$G^{(1)}(x) = \sum_n x^n P^{(1)}_n$$ and its generalization to  $s$
generations. For example, $P^{(2)}_n$ is the chance to have $n$
grandchildren, while $G^{(2)}(x)$ is the corresponding generating
function. The successive generating functions are given by
\begin{equation}
G^{(s)}(x) = G^{(1)}(G^{(s-1)}(x)). \label{eq1}
\end{equation}
By definition, $G^{(s)}(0) = P^{(s)}_0$ is the probability that the
lineage originated from the founder went extinct by the $s$th
generation, and $G^{(s+1)}(0)-G^{(s)}(0) \approx dG^{(s)}(0)/ds$
 determines the chance of extinction at the $s$
generation. After many generations the chance of extinction is
almost one, so $G^{(s)}(0) = 1-\delta_s$, $\delta_s  \ll 1$.  Eq.
(\ref{eq1}) implies that $$1-\delta_{s+1} = \sum_n P^{(1)}_n
(1-\delta_s)^n \approx \sum_n P^{(1)}_n \left(
1-n\delta_s+n(n+1)\delta^2_s/2 \right),$$ which leads to the
recursive equation $$\delta_{s+1} = \delta_{s} - \sigma^2
\delta^2_{s}/2.$$ Taking the continuum limit we then obtain that the
survival probability is determined by the differential equation
$$\dot{\delta}(t) \sim -\sigma^2 \delta^2(t)/2.$$ This result implies
that the long-time decay of the survival probability goes like
$\delta(t) \approx 2/(\sigma^2 t)$, so the statistics of species
lifetime admits a $1/t^2$ tail \cite{pigolotti2005}.

\section{Appendix  B: Survival probability and age-abundance relationships in
a model with both demographic and environmental noise}

\subsection{I. Neutral dynamics with environmental noise: The geometric neutral process}

In this section we derive our main results, implementing a model
with a geometric distribution of offspring. As explained in the main
text, two cases are considered here: a logarithmically balanced
noise and an arithmetically balanced noise.

In a neutral model with a geometric distribution but  without any
stochasticity, the chance of an individual to produce $n$ offspring
is $P_n = 1/2^{n+1}$ and the corresponding generating function for a
single generation is $$ G^{({1})}(x) = \sum_n x^n \ P_n =
\frac{1}{2-x}.$$ What makes this model easy to handle is the
convenient structure of the generating function. The generating
function for the population after $s$ generations is obtained by
iterating $G^{(1)}$, and in this specific case the answer is
immediate \cite{steffensen1933},
\begin{eqnarray} \label{eq3}
G^{(s)}(x) &\equiv& \sum_{n=0}^{\infty}x^n P^{(s)}_n =
G(G(G...{\rm{s \  times}} \ (G(x)))) =  \frac{s-(s-1)x}{(s+1)-sx}.
\end{eqnarray}
Thus the distribution, apart from $P_0$, remains geometric. From
this one can easily derive the results given above for a purely
demographic model for this special case of geometrically distributed
births.

To construct a model combining both demographic and environmental
noise, we consider a discrete time dynamics, where for convenience
we choose the time step to equal the generation time, so that all
individuals of the $s$th generation reproduce and then
simultaneously pass from the scene.

Now let us consider the two  random processes defined in the main
text. The probability for each individual at the $s$-th generation
to produce $n$ offspring conditioned on an environmental noise
determined $\gamma_s$ is
\begin{eqnarray} \label{eq3}
P(n|\gamma_s) = \frac{e^{\gamma_sn}} {(e^{\gamma_s}+1)^{n+1}} \qquad
\rm{case \  A} \nonumber \\
P(n|\gamma_s) = \frac{(1+\gamma_s)^n} {(2+\gamma_s)^{n+1}} \qquad
\rm{case \  B},
\end{eqnarray}
where \textbf{case A} is the logarithmically balanced noise and
\textbf{case B} corresponds to arithmetically balanced noise.

When $\gamma_s = 0$ for any $s$ one obtains, of course, a purely
demographic process. For nonzero  $\gamma_s$  the fitness (or the
deterministic growth rate) of the population fluctuates,
 and $D \equiv \overline{\gamma^2}/2$ characterizes the strength of the environmental noise. The model is neutral in the
 sense that $\gamma_s$   is distributed
identically for all species and so considered over long time scales,
all species are demographically equivalent.

The  generating functions in the two cases are then,
\begin{eqnarray}
G^{(1)}(x|\gamma_s) = \frac{1}{1+e^{\gamma_s}-xe^{\gamma_s}}, \qquad
\rm{case \ A}; \nonumber \\
G^{(1)}(x|\gamma_s) = \frac{1}{2+\gamma-x(1+\gamma)}, \qquad
\rm{case \ B.} \label{g1}
\end{eqnarray}

Using the recurrence relation  Eq. (\ref{eq1}) and the offspring
generating function, Eq. (\ref{g1}), one can
obtain~\cite{steffensen1933} the general form of
$G^{(s)}(x|\gamma_s) $  as
\begin{small}
\begin{equation} \label{eq4}
G^{(s)}(x|\gamma_s)  = \frac{a_s+b_s\cdot x}{c_s+d_s\cdot x},
\end{equation}
\end{small}
\noindent where the values of the constants
\begin{small}$a_s,b_s,c_s$\end{small} and
\begin{small}$d_s$\end{small} satisfy [for a log-balanced noise (case A)] the recurrence
relation
\begin{small}
\begin{equation} \label{eq5}
\left[ \begin{array}{c} a_{s+1} \\ c_{s+1} \\ b_{s+1} \\
d_{s+1}
\end{array} \right] =
\begin{bmatrix} 0 & 1 & 0 &0  \\ -e^{\gamma_{s+1}} & e^{\gamma_{s+1}} + 1  & 0 & 0 \\ 0 & 0 & 0 &  1  \\ 0 & 0 & -e^{\gamma_{s+1}} & e^{\gamma_{s+1}} + 1  \end{bmatrix}  \left[
\begin{array}{c} a_{s} \\ c_{s} \\ b_{s} \\ d_{s} \end{array}
\right]
\end{equation}
\end{small}
\noindent with the initial conditions  \begin{small} $a_1 = 1, \ b_1
= 0, \ c_1 = e^{\gamma_1}+1$\end{small} and \begin{small}$d_1 =
-e^{\gamma_1}$\end{small}. The corresponding equations in case B are
obtained by replacing $e^\gamma$ by $1+\gamma$ in all the above
expressions.

 The chance of survival until the $s$th generation
\begin{small}$\delta_s \equiv 1-G^{s}(0)$\end{small} is simply
\begin{small}$$\delta_s = 1-a_s/c_s.$$\end{small}
Accordingly, the survival probability satisfies the stochastic
recursion relations:
\begin{eqnarray}\label{eq:S_s}
\delta_{s+1} = \frac{\delta_s}{e^{-\gamma_s}+\delta_s}, \qquad \rm{case \ A}; \nonumber \\
\delta_{s+1} = \frac{(1+\gamma) \delta_s}{1+(1+\gamma) \delta_s},
\qquad \rm{case \ B.}
\end{eqnarray}

When $\delta_s \ll 1$, it satisfies $\delta_{s+1} = \delta_s
e^{\gamma_s} -\delta^2_s$ in case A and $\delta_{s+1} = (1+\gamma)
\delta_s  -\delta^2_s$ in case B.  One has to be careful in
translating the equation for case  A to a stochastic differential
equation, as $exp(\gamma_s)$ has a non-zero expectation value,
namely $\cosh \sqrt{2D}$.  Thus, in terms of $\delta$ there is an
extra term driving $\delta$ to larger values.  Taking this into
account, in the weak environmental noise limit, $\gamma_s \ll 1$,
one gets the stochastic differential equations that were quoted in
the main text:

\begin{eqnarray} \label{res}
 \frac{d \delta(t)}{dt} =  \gamma_t \delta(t)  + D \delta(t) - \delta^2(t), \qquad \rm{case \ A}; \nonumber \\
  \frac{d \delta(t)}{dt} =  \gamma_t \delta(t)   - \delta^2(t), \qquad \rm{case \ B.}
\end{eqnarray}

\subsection{II. Long-time asymptotics}

The Langevin equations (\ref{res}) should be interpreted in the Ito
sense, since they were derived as the white noise, continuous time
limit of a non-overlapping generation model that assumes zero
relaxation time, i.e., that the population follows the instantaneous
growth rate determined by $\gamma$
\cite{kupferman2004ito,suweis2011prescription}. When the relaxation
time of the population or the community is finite (i.e., when the
demographic rates respond slowly to the changing environment, with
respect to the noise correlation time) the effective strength of the
environmental noise amplitude decreases, but the interpretation is
still Ito. Accordingly,  in the logarithmic space $y = ln(\delta)$,
Eqs. (\ref{res}) takes the form:
\begin{eqnarray} \label{eqy}
 \frac{d y}{ds} =  \gamma(s) - \exp(y), \qquad \rm{case \ A}; \nonumber
 \\
 \frac{d y}{ds} =  \gamma(s) - D - \exp(y), \qquad \rm{case \
 B}.
\end{eqnarray}
It is clear, now, that as $\delta \to 0$, i.e., $y \to -\infty$, in
case A the system  performs an unbiased random walk in the log
space, hence in the long run one expects that the probability of a
taxon to have a lifetime $t$  will behave like $t^{-3/2}$, i.e.,
that $\delta(t) \sim t^{-1/2}$. In case B, on the other hand, the
random walk is biased to the left, and the survival probability
should decrease exponentially in time.

\begin{figure}
\begin{center}
\includegraphics[width=7cm]{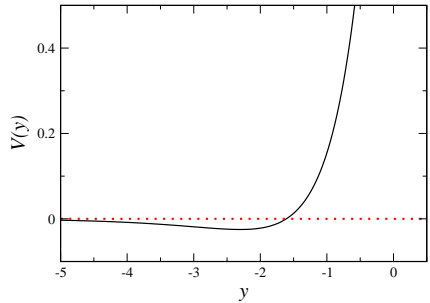}
\vspace{-0.8cm}
\end{center}
\caption{The Morse potential $V(y)= \exp(2y)/4D \ - \ \exp(y)/2$ for
 $D=0.1$.} \label{morseDp1}
\end{figure}

 To demonstrate that, we notice that the corresponding Fokker-Planck Equations
 are
\begin{eqnarray}
 \partial_s P(y,s) = D \partial^2_y P(y,s) + \partial_y \left( \exp(y) P(y,s) \right), \qquad \rm{case \ A;} \nonumber \\
 \partial_s P(y,s) = D \partial^2_y P(y,s) + \partial_y \left([D+ \exp(y)] P(y,s) \right),  \qquad \rm{case \ B}.
\end{eqnarray}
 With the substitutions $P(y,s) = \exp(-e^y/2D)
\psi(y,s)$ (A) and  $P(y,s) = \exp(-y/2-e^y/2D) \psi(y,s)$ (B) one
gets
\begin{eqnarray}
 \dot{\psi}(y,s) = D \psi'' +  \left( \frac{\exp(y)}{2}- \frac{\exp(2y)}{4D}\right)
 \psi,  \qquad \rm{case \ A}; \nonumber \\
\dot{\psi}(y,s) = D \psi'' -  \left( \frac{\exp(2y)}{4D} +
\frac{D}{4} \right) \psi, \qquad \rm{case \ B.}
 \label{FPmorse}
\end{eqnarray}
These  are  Schrodinger equations in imaginary time with an
exponentially decaying potential (case B) and a Morse potential
(case A, see Fig. \ref{morseDp1}).

\begin{figure}
\begin{center}
\includegraphics[width=7cm]{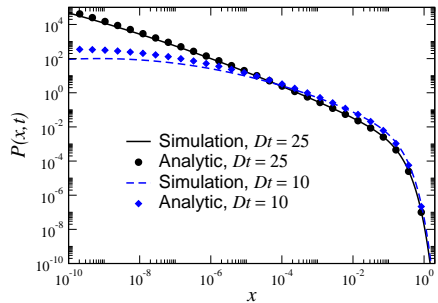}
\vspace{-0.8cm}
\end{center}
\caption{Numerical simulation of the Fokker-Planck Eq. (\ref{res}),
compared to the analytic approximation, Eq. (\ref{res2}), for
$D=0.1$, $Dt=10$ and $Dt=25$.} \label{MorsePx}
\end{figure}

A wavepacket which is initially localized at small negative  values
of $y$ (corresponding to $\delta\sim 1$) will eventually reach the
region of large negative $y$ ($\delta\rightarrow0$), where the
potential is negligible and the motion is almost purely diffusive.
This is clearer in case B, where the long-term behavior of $\psi$
(neglecting the exponential term) is,
\begin{equation}
\psi(y,t) \approx \frac{e^{-\frac{y^2}{4Dt}}e^{-Dt/4}}{t}
\end{equation}
so the wavefunction diffuses in the log space, but this diffusion is
superimposed of an exponential decay $\exp(-Dt/4)$ (we have replaced
the generation parameter $s$ by time $t$, as in the long-term the
changes of $\delta$ over a single generation are small). For case A
an asymptotic matching approach  leads to the uniform approximation
\begin{equation} \label{res2}
  P(\delta,t) \approx  \frac{1}{\delta\sqrt{\pi Dt}} e^{-(ln(\delta/D)+\gamma_E)^2/4Dt}
  e^{-\delta/D}.
\end{equation}
Where $\gamma_E$ is the Euler constant. This approximate solution is
shown in Fig. \ref{MorsePx}, together with a direct numerical
solution of the Fokker-Planck Eq. (\ref{res}).  We see that the
agreement is very good, and improves with time. At  long times, the
expectation value of $\delta$, $\overline{\delta}  \approx
\sqrt{D/\pi}$, so that the extinction rate has a  $t^{-3/2}$ tail.
For case B we have,
\begin{equation}
P(\delta,t) \approx \frac{1}{\delta \sqrt{\pi D t}} e^{\frac{(ln
\delta +D)}{4Dt}}
\end{equation}
so that the log-normal distribution moves toward negative values of
$ln(\delta)$ at a constant rate, leading to a typical $\delta$ which
decays exponentially in time.

\subsection{III: Abundance-age relationship}

We now turn to analyze the effect of environmental noise on the
abundance-age relationship in neutral theory. As explained in the
main text, there are two ingredients that determine the
age-abundance relationships: the chance of a species to reach
abundance $N$ after $s$ generations, and its  chance of survival.
The ratio between these factors gives $N_{surv}$, the average size
of a surviving species. The chance of survival was calculated above,
and now we will calculate the average abundance.

In case B the result is quite trivial: since  the demographic and
the environmental stochasticity are both  balanced in the abundance
space, the average abundance of a species is fixed. Accordingly, we
perform the analysis here for case A, and will explain at the end
how the results relate to our prediction for case B.

From Eq. (\ref{eq4}), we can see that the width of the geometric
distribution of the abundance at time $s$ is given by
\begin{small}$\zeta(s) = -1/\ln(-d_s/c_s)$\end{small}. Note that
\begin{small}$d_s$\end{small} and \begin{small}$c_s$\end{small}
satisfy the same recurrence relations, and the only difference
between them is the initial conditions. This allows us to consider
only \begin{small}$d_s$\end{small} and then to carry over the
results to \begin{small}$c_s$\end{small} and hence calculate their
ratio. From (\ref{eq5}) it is clear that
\begin{small}$d_s$\end{small} satisfies
\begin{small}
\begin{equation} \label{rec}
d_{s+1} - d_{s}  = e^{\gamma_s}(d_{s} - d_{s-1}).
\end{equation}
\end{small}
The solution to this is easily verified to be
\begin{small}
\begin{equation}
d_s =  d_1 \sum_{k=1}^{s} \prod_{j=2}^{k} e^{\gamma_j} .
\end{equation}
\end{small}
Given $d_1=-e^{\gamma_1}$, we can write this as
\begin{small}
\begin{equation}
d_s =  - \sum_{k=1}^{s} \prod_{j=1}^{k} e^{\gamma_j} .
\end{equation}
\end{small}
 As explained above, the equations for $c$ is the same, except that $c_1=1+e^{\gamma_1}$, so that
\begin{small}
\begin{equation}
c_s = \sum_{k=0}^{s} \prod_{j=1}^{k} e^{\gamma_j} .
\end{equation}
\end{small}
Thus, $d_s = - (c_s - 1)$ and so $1/\zeta = -\ln(-d_s/c_s)=-\ln(1 -
1/c_s)$.   It should first be noted that for $s\gg 1$, $D \ll 1$,
$c_s$ is large.  For the case of binary noise, for example,
$\gamma=\pm \gamma_0$, the minimum possible value of $c_s$ is for
the case where $\gamma_s$ is always negative, in which case $c_s
\approx 1/\gamma_0$. Thus, $\zeta \approx c_s$, so that
\begin{equation}
\overline{\zeta} \approx \sum_{k=0}^s \overline{\prod_{j=1}^k
e^{\gamma_j}} = \sum_{k=0}^s e^{Dk/2} =
\frac{2}{D}\left(e^{Ds/2}-1\right) \label{eqwidth}
\end{equation}
So we see that $\overline{c_s}$ is exponentially large for large
$s$.

\begin{figure*}
\begin{center}
\includegraphics[width=10cm]{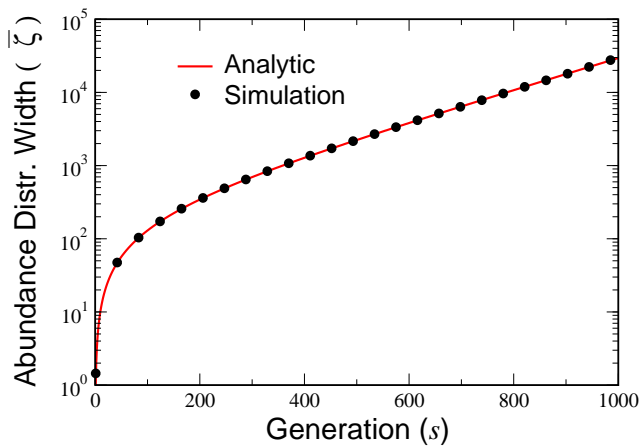}
\end{center}
\vspace*{-0.8cm} \caption{A graph of the ensemble-averaged width
$\overline{\zeta(s)}$ calculated by averaging $\zeta$ as computed
from the analytic generating function Eq. (\ref{eq5}) over $10^6$
realizations of binary noise $\gamma=\pm \gamma_0$, with
$\gamma_0=0.1$.  Also show in the analytic formula, Eq.
(\ref{eqwidth}).} \label{figwidth}

\end{figure*}

For case B, as explained, one should replace $e^\gamma$ by
$1+\gamma$, and the trivial result is a linear growth, $\zeta \sim
t$.

The result (\ref{eqwidth}) has two limits: when $s\ll 1/D$, the
environmental noise is negligible and the average abundance of a
species (conditioned on its presence in the community) grows
linearly in time. However, at long times the environmental noise
controls the system and the typical size of surviving species grows
exponentially with $s$. In this regime the typical timescale (in
generations) needed for a species to reach abundance $N$ will scale
like $log(N)/D$, as opposed to the linear $N$ scaling for pure
demographic noise.

\subsection{IV: Average vs. typical age-abundance relationships in case A}
Our result for case A, Eq. (\ref{eqwidth}), might appear at first
glance to  be in contradiction with the results proved by Geiger and
collaborators \cite{geiger2003limit}. These authors show that, in
case A
\begin{equation} \label{ggr}
lim_{s \to \infty} \frac{log[E(N_{surv})]}{s} = 0
\end{equation}
and decided that this log-balanced scenario is ``critical'', as
opposed to case B (or, in general, any noise which gives a negative
bias in the log-space) which is subcritical.

To explain this apparent contradiction, let us stick to the method
used in \cite{geiger2003limit}, averaging first over demographic
noise. In the log-abundance space, what one has after taking this
average is a random walker that starts at zero (one individual) and
moves randomly with no bias until $s$. Clearly, the population at
$s$, conditioned on survival and averaged over demographic noise, is
$exp(y)$, where $y$ is the location of a random walker that starts
at $s=0$ and  subsequently never crosses zero, since otherwise  it
goes extinct. Such a constrained random walk is known as a meander.
To calculate the probability that the meander is at $y$ at time $s$
one solves for  a RW that start at $x_o$ using images, taking $x_0$
to zero at the end. The result is
\begin{equation}
P(y,s) = A \frac{ye^{-y^2/4Ds}}{s^{3/2}}.
\end{equation}
Accordingly,  the maximum likelihood for the position of the meander
at s is (for large s),
\begin{equation}
Y_{ML} = \sqrt{2Ds}.
\end{equation}
This implies that the typical size of a population at $s$,
conditioned on non-extinction is,
 \begin{equation}
N^{typ}_{surv} \approx e^y =  \exp(\sqrt{2Ds}).
\end{equation}
On the other hand, the average value of $ N_{surv} $ will be,
 \begin{equation}
N_{surv} \approx \int dy \  e^y \frac{ye^{-y^2/4Ds}}{s^{3/2}}
\approx e^{Dt}.
\end{equation}
Note that the result for the average comes from the peak of the
integrand at $y^*=2Ds$ (this is a Laplace integral) and $P(y^*)$ is
exponentially small as $s \to \infty$, so this result is consistent
with \cite{geiger2003limit} since the average comes from
exponentially rare events (with probability one this will not be the
case for any specific history). On the other hand, $P(y_{ML})$ is
${\cal O}(1/t)$, which is not  negligible.  However, the typical
growth is subexponential, and clearly agrees with (\ref{ggr}).

\section{Appendix C: Universality}

Throughout section B we have studied the geometric neutral process
because of its convenient properties. Of course, any result that
depends on the specific properties of a certain distribution cannot
be relevant to the generic case, where the distribution of number of
descendents per individual is, in most cases, unknown and there is
no reason to believe that it belongs to any particular simple
distribution.

To illustrate the generality of our results, let us consider the
survival probability $\delta_s$. In the generic case one may define
a probability distribution function $P_n(s)$, the chance of an
individual to produce $n$ offspring during the $s$ generation, and
of course the two important summary statistics that characterize
this distribution are its mean (in case A, ${\bar n} =
exp(\gamma_s)$, in case B, ${\bar n} = \gamma_s$ and its variance.
Here we assume that the environmental noise is weak so one can
neglect the variance fluctuations and $var(n)=\sigma^2$, where
$\sigma^2$ is the variance of the purely demographic model
($\sigma^2 = 1$ for Poisson distribution with average one, $\sigma^2
= 2$ for the geometric distribution and so on).

The generating function recursion relation implies that
\begin{equation} \label{c1}
G^{(s)}(x) = G^{(1)}_s(G^{(s-1)}(x)).
\end{equation}
In the long time limit the chance to survive is small, so
$G^{(s)}(x) = 1-\delta_s$. Plugging this into (\ref{c1})  one gets:
\begin{eqnarray} \label{c2}
1- \delta_{s} = \sum_n P_n(s) \left(1-\delta_{s-1} \right)^n \approx
\sum_n P_n(s) \left(1-n \delta_{s-1}+\frac{n^2-n}{2} \delta^2_{s-1}
\right).
\end{eqnarray}
Accordingly:
\begin{eqnarray} \label{c3}
{\dot \delta} \approx \delta_{s} - \delta_{s-1} &=& (e^{\gamma_s}-1) \delta - \frac{\sigma^2}{2} \delta^2 \qquad \rm{case \ A} \nonumber \\
 {\dot \delta} &=& \gamma_s \delta - \frac{\sigma^2}{2} \delta^2   \qquad \rm{case \ B}.
\end{eqnarray}
The second equation is what appears in Eq. 4 of the main text. In
case A one needs to expand the exponent in $\gamma$ and to replace
$\gamma^2/2$ by $D$ to obtain the  equation presented in the main
text.

\end{document}